\documentclass[12pt,draftclsnofoot,onecolumn]{IEEEtran}
\usepackage{graphicx,cite,epsfig,amssymb,amsmath}

\begin{document}
\markboth{IEEE Transaction on WIRELESS COMMUNICATIONS, Vol. XX,
No. Y, Month 2011} {}

\title{\mbox{}\vspace{1.00cm}\\
\textsc{Interference Mitigation for Cognitive Radio MIMO Systems Based on Practical Precoding} \vspace{1.5cm}}

\author{\normalsize
Zengmao Chen$^{1}$, Cheng-Xiang Wang$^{1}$, Xuemin Hong$^{1}$, John Thompson$^{2}$\\
Sergiy A. Vorobyov$^{3}$, Feng Zhao$^{4}$, Hailin Xiao$^{5}$, and  Xiaohu Ge$^{6}$\\
\vspace{0.7cm}
$^{1}$ Joint Research Institute for Signal and Image Processing\\
School of Engineering \& Physical Sciences\\
Heriot-Watt University, Edinburgh, EH14 4AS, UK.\\
Email:\{zc34, cheng-xiang.wang, x.hong\}@hw.ac.uk\\
\vspace{0.3cm}
$^{2}$Joint Research Institute for Signal and Image Processing\\
Institute for Digital Communications\\
University of Edinburgh,Edinburgh, EH9 3JL, UK.  \\
Email: john.thompson@ed.ac.uk\\
\vspace{0.3cm}
$^{3}$ Department of Electrical and Computer Engineering\\
University of Alberta, Edmonton, AB, T6G 2V4, Canada. \\
Email: vorobyov@ece.ualberta.ca\\
\vspace{0.3cm}
$^{4}$Department of Science and Technology\\
Guilin University of Electronic Technology, Guilin 541004, China.\\
Email: zhaofeng@guet.edu.cn\\
\vspace{0.3cm}
$^{5}$School of Information and Communication\\
Guilin University of Electronic Technology, Guilin 541004, China.\\
Email: xhl\_xiaohailin@yahoo.com.cn\\
\vspace{0.3cm}
$^{6}$Department of Electronics and Information Engineering\\
Huazhong University of Science and Technology, Wuhan 430074, China.\\
Email: xhge@mail.hust.edu.cn\\
}
\date{\today}
\renewcommand{\baselinestretch}{1.2}
\thispagestyle{empty} 
\maketitle
\thispagestyle{empty}
\newpage
\setcounter{page}{0}

\begin{abstract}

In this paper, we propose two subspace-projection-based precoding schemes, namely, full-projection (FP)- and partial-projection (PP)-based precoding, for a cognitive radio multiple-input multiple-output (CR-MIMO) network to mitigate its interference to a primary time-division-duplexing (TDD) system. The proposed precoding schemes are capable of estimating interference channels between CR and primary networks, and incorporating the interference from the primary to the CR system into CR precoding via a novel sensing approach. Then, the CR performance and resulting interference of the proposed precoding schemes are analyzed and evaluated. By fully projecting the CR transmission onto a null space of the interference channels, the FP-based precoding scheme can effectively avoid interfering the primary system with boosted CR throughput. While, the PP-based scheme is able to further improve the CR throughput by partially projecting its transmission onto the null space.\\


\end{abstract}
\newpage
\IEEEpeerreviewmaketitle

\section{Introduction}

A cognitive radio (CR) \cite{cr1}--\!\!\cite{cr4} system may coexist with a primary network on an either interference-free or interference-tolerant basis \cite{cr3}. For the former case, the CR system only exploits the unused spectra of the primary network. While, for the latter case, the CR system is allowed to share the spectra assigned to primary network under the condition that it must not impose detrimental interference on the primary network. Therefore, the interference from the CR network to the primary system should be carefully managed and cancelled in order to protect the operation of the primary system. Various interference mitigation (IM) techniques applicable to CR networks have been reported in \cite{ic_magazine}. As for multiple-antenna CR networks, transmit beamforming (for single-data-stream transmission) \cite{zhoujun}--\!\!\cite{kit2} or precoding (for multiple-data-stream transmission) \cite{luca}--\!\!\cite{cr_mimo} is an effective approach to proactively cancel interference from CR transmitters to the primary network. On one hand, it steers the CR transmission to avoid interfering with the primary network. On the other hand, it exploits the diversity or the multiplexing gain of the multiple-antenna CR system to enhance the reliability or efficiency of the CR system. 

However, in the works \cite{zhoujun}--\!\!\cite{cr_mimo}, perfect or partial CSI of CR interference channels to primary network (CR-primary interference channels) is required at the CR transmitter side to guarantee no/constrained interference to the primary system. Therefore, extra signaling between primary and CR networks is inevitable to obtain the CSI, which jeopardizes the applicability of these beamforming and precoding schemes. A more practical precoding scheme  - sensing-projection (SP)-based precoding, which learns the CSI using subspace estimation \cite{subspace} and does not require a priori CSI, has been proposed for a CR multiple-input multiple-output (MIMO) transmitter-receiver pair coexisting with a primary time-division-duplexing (TDD) system in \cite{ns_precoding, practical_bf}. However, such precoding scheme does not account for the interference from primary transmitters to the CR receiver (primary-CR interference), 
which leads to a CR throughput loss. In \cite{yi} and \cite{gao}, it is proposed to remove the primary-CR interference at the CR receiver via null-space receiver beamforming, which sacrifices the CR throughput as well. Moreover, the CR network in \cite{yi, gao} has to work in a TDD mode aligned with the primary system in order to facilitate the null-space receiver
beamforming.

In this paper, two enhanced SP-based precoding schemes, namely, full-projection (FP)- and partial-projection (PP)-based precoding, are proposed for CR MIMO systems by incorporating the primary-CR interference. As the name suggests, the FP-based scheme nulls the CR transmission by fully projecting the transmission onto the estimated null space of the CR-primary interference channels. Instead of removing the primary-CR interference using null-space receiver beamforming, the proposed precoding schemes account for the primary-CR interference via sensing. This, on one hand, improves the CR throughput, and on the other hand, introduces more flexibility into the CR deployment, i.e., the CR network does not have to work in a TDD mode as in \cite{yi, gao}.
The PP-based precoding can further improve the CR throughput by projecting the CR transmission onto a subspace that partially spans the estimated null space of the CR-primary interference channels. As a result, the CR throughput is further improved at the cost of introducing extra interference to the primary network.

The remainder of this paper is organised as follows. The system model is given in Section~II. The principle of the SP-based precoding is introduced in Section~III. The new precoding schemes are proposed in Section IV. The performance of the proposed precoding schemes is evaluated in Section~V. Finally, we conclude the paper in Section~VI.



\section{System Model and Problem Formulation}

We consider a CR system shown in Fig.~1, where a CR transmitter-receiver pair shares the same spectrum with a primary TDD network. Multiple antennas are mounted at the CR nodes and possibly at each of the primary users. The CR transmitter, CR receiver, primary base station (BS) and the $k$th primary user are equipped with $M_t$, $M_r$, $M_{bs}$ and $M_k$ ($k=1,\cdots,K$) antennas, respectively. Block-fading channels are assumed for the primary and CR systems.

For a narrowband transmission, the received symbol at the CR receiver can be expressed as
\begin{equation}\label{eq1}
\vspace{-0.4cm}
\mathbf{y}=\mathbf{H}\mathbf{F}\mathbf{s} + \mathbf{n} +\mathbf{z}
\end{equation}
where $\mathbf{y}\in \mathbb{C}^{M_r \times 1}$ is the received signal vector at the CR receiver, $\mathbf{s}\in \mathbb{C}^{M_t \times 1}$ and $\mathbf{F}\in \mathbb{C}^{M_t \times M_t}$ are the transmit information vector with $\mathbb{E}\{\mathbf{s}\mathbf{s}^H\}=\mathbf{I}$ and precoding matrix of the CR transmitter, respectively, $\mathbf{H}\in \mathbb{C}^{M_r\times M_t}$ is the channel matrix from the CR transmitter to CR receiver, whose elements are independent and identically distributed (i.i.d.) complex Gaussian random variables with zero mean and variance $\sigma^2_H$, and $\mathbf{n}\in \mathbb{C}^{M_r \times 1}$ stands for the additive white Gaussian noise (AWGN) vector with zero mean and covariance matrix $\mathbb{E}\{\mathbf{n}\mathbf{n}^H\}=\sigma^2_n\mathbf{I}$. Moreover, $\mathbf{z}\in \mathbb{C}^{M_r \times 1}$ denotes the interference from the primary network to CR receiver. It can be expressed as
\begin{equation}\label{eq2}
\mathbf{z} = \left\{
\begin{array}{ll}
\mathbf{H}_{\rm{ur}}\mathbf{x}_{u},\ \ \ \ \ \rm{during\ primary\ uplink}\\
\mathbf{H}_{\rm{dr}}\mathbf{x}_{d},\ \ \ \ \ \rm{during\ primary\ downlink}.\\
\end{array} \right.
\end{equation}

The overall CR system is shown in Fig.~1. The matrices $\mathbf{H}_{\rm{ur}}\in \mathbb{C}^{M_r \times \sum^K_{k=1}M_k}$ in \eqref{eq2} and $\mathbf{H}_{\rm{ut}}\in \mathbb{C}^{ M_t \times \sum^K_{k=1}M_k}$ (see Fig.~1) represent the interference channels from all the $K$ active primary users to CR receiver and to CR transmitter, respectively, during primary uplink. Similarly, $\mathbf{H}_{\rm{dr}}\in \mathbb{C}^{ M_r \times M_{bs}}$ in \eqref{eq2} together with $\mathbf{H}_{\rm{dt}}\in \mathbb{C}^{ M_t \times M_{bs}}$ (see Fig.~1) stand for the interference matrices from the primary BS to CR receiver and to CR transmitter during primary downlink. All these interference matrices ($\mathbf{H}_{\rm{ur}}$, $\mathbf{H}_{\rm{ut}}$, $\mathbf{H}_{\rm{dr}}$ and $\mathbf{H}_{\rm{dt}}$) have i.i.d. complex Gaussian random elements with zero mean and covariances $\sigma^2_{\rm{ur}}$, $\sigma^2_{\rm{ut}}$, $\sigma^2_{\rm{dr}}$ and $\sigma^2_{\rm{dt}}$, respectively. Moreover, $\mathbf{x}_{u}\in \mathbb{C}^{\sum^K_{k=1}M_k \times 1}$ and $\mathbf{x}_{d}\in \mathbb{C}^{M_{bs}\times 1}$ are the transmitted signal vectors of all the $K$ primary users and primary BS, respectively. We define the covariance matrix of the interference in \eqref{eq2} as $\mathbf{Z}\triangleq\mathbb{E}\{\mathbf{z}\mathbf{z}^H\}$.
\section{Principle of SP-based Precoding}

The precoding problem for CR transmission can be expressed as the following optimisation problem \cite{rui}
\vspace{-0.5cm}
\begin{align}\label{eq3}
&\max_{\mathbf{F}}\ \  \rm{log}_2 \rm{det}\left( \mathbf{I} + \frac{\mathbf{H}\mathbf{F}\mathbf{F}^H\mathbf{H}^H}{\sigma^2_{n}}\right) \\ \label{c2}
&{\rm subject\ to}\ \ \  \mathtt{Tr}\{\mathbf{F}\mathbf{F}^H\}\leq P_{cr} \\ \label{c3}
&\qquad\qquad\ \ \ \  \mathtt{Tr}\{\mathbf{G}_k\mathbf{F}\mathbf{F}^H\mathbf{G}^H_k \}\leq \Gamma_k \ \ k=1,\ldots,L \ .
\end{align}
In \eqref{c3}, $\mathbf{G}_k \in \mathbb{C}^{M_k\times M_t}$ is the channel matrix from the CR transmitter to the $k$th primary user. Thus, the channel matrix from the CR transmitter to all primary users becomes $\mathbf{H}^H_{\rm{ut}}\!=\![\mathbf{G}^T_1,\cdots,\mathbf{G}^T_K]^T$, where channel reciprocity is assumed. The constraints on the CR transmission power and the maximum allowed interference perceived at each primary user are given by \eqref{c2} and \eqref{c3}, respectively.

The projected channel singular value decomposition (SVD) or P-SVD precoding has been proposed in \cite{rui} as a suboptimal solution for the optimisation problem \eqref{eq3}--\eqref{c3}. It can be expressed as
\vspace{-0.3cm}
\begin{equation}\label{eq5}
\mathbf{F}=\mathbf{U}_{\perp}\left[(\mu \mathbf{I} -\mathbf{\Lambda}^{-1}_{\perp})^{+}\right]^{\frac{1}{2}}
\end{equation}
where $(\cdot)^+=\rm{max}(0,\cdot)$, $\mu$ denotes the power level for a water-filling (WF) algorithm, and $\mathbf{U}_{\perp}$ and $\mathbf{\Lambda}_{\perp}$ originate from the SVD of the effective CR channel matrix $\mathbf{H}_{\perp}$
\begin{equation}\label{eq4}
\mathbf{H}_{\perp}\triangleq \mathbf{H}(\mathbf{I}-\mathbf{U}_G\mathbf{U}^H_G).
\end{equation}
Its SVD is expressed as $\mathbf{H}_{\perp}=\mathbf{V}_{\perp}\mathbf{\Lambda}^{1/2}_{\perp}\mathbf{U}^H_{\perp}$, and $\mathbf{U}_G$ in \eqref{eq4} is from another SVD $\mathbf{H}^H_{\rm{ut}}=\mathbf{V}_{G}\mathbf{\Lambda}^{1/2}_{G}\mathbf{U}^H_{G}$, which is estimated via sensing in the SP precoding \cite{ns_precoding, practical_bf} as shown in Fig.~2.

By analogy with the multiple signal classification technique \cite{subspace}, the signal covariance matrix is decomposed into signal and noise subspaces to estimate $\mathbf{U}_G$, which can be mathematically\! expressed\! as
\vspace{-0.4cm}
\begin{align}\label{eq7}
\hat{\mathbf{R}}_{\rm{ut}}&=\frac{1}{L_S}\sum^{L_S}_{i=1}\mathbf{r}_{\rm{ut}}(i) \mathbf{r}^H_{\rm{ut}}(i)\\ \label{svd}
&=\hat{\mathbf{U}}\hat{\mathbf{\Lambda}}\hat{\mathbf{U}}^H \\ \label{gn}
&=\hat{\mathbf{U}}_G\hat{\mathbf{\Lambda}}_G\hat{\mathbf{U}}^H_G + \hat{\mathbf{U}}_n\hat{\mathbf{\Lambda}}_n\hat{\mathbf{U}}^H_n.
\end{align}
In \eqref{eq7}, $\mathbf{r}_{\rm{ut}}(i)=\mathbf{H}_{\rm{ut}}\mathbf{x}_{u}(i) + \mathbf{n}(i)$ is the $i$th received symbol at the CR transmitter, and $\hat{\mathbf{R}}_{\rm{ut}}$ denotes the average covariance matrix of the received symbols. An eigenvalue decomposition is then performed on $\hat{\mathbf{R}}_{\rm{ut}}$ in \eqref{svd}, where $\hat{\mathbf{\Lambda}}=\rm{diag}(\lambda_1,\cdots,\lambda_{M_t})$ is a diagonal matrix with descendingly ordered eigenvalues of $\hat{\mathbf{R}}_{\rm{ut}}$ and $\hat{\mathbf{U}}\in \mathbb{C}^{M_t\times M_t}$ contains the corresponding eigenvectors. The matrix $\hat{\mathbf{R}}_{\rm{ut}}$ is further decomposed into interference and noise components in \eqref{gn} with $\hat{\mathbf{U}}_G$ and $\hat{\mathbf{U}}_n$ being the first $K_p=\rm{rank}(\mathbf{H}_{\rm{ut}})$ and the remaining $(M_t\!-\!K_p)$ columns of $\hat{\mathbf{U}}$, respectively, and $\hat{\mathbf{\Lambda}}_G$ and $\hat{\mathbf{\Lambda}}_n$ being their corresponding eigenvalue matrices.



\section{New Precoding Schemes}

In this section,
we elaborate the CR precoding during the primary downlink. A similar precoding for the primary uplink can be easily obtained, which is ignored here for brevity. When incorporating the primary-CR interference, the precoding problem for the CR transmitter during the primary downlink can be expressed as the following optimisation problem  
\begin{align}\label{eq9}
&\max_{\mathbf{F}}\ \  \rm{log}_2 \rm{det}\left( \mathbf{I} + \frac{\mathbf{H}\mathbf{F}\mathbf{F}^H\mathbf{H}^H}{\mathbf{Z} + \sigma^2_{n}\mathbf{I}}\right) \\ \label{cc2}
&{\rm subject\ to}\ \ \ \mathtt{Tr}\{\mathbf{F}\mathbf{F}^H\}\leq P_{cr} \\ \label{cc3}
&\qquad\qquad\ \ \ \  \mathtt{Tr}\{\mathbf{G}_k\mathbf{F}\mathbf{F}^H\mathbf{G}^H_k \}\leq \Gamma_k, \ \ k=1,\ldots,L.
\end{align}
\vspace{-0.3cm}
Then, the precoding matrix for CR transmission during the downlink can be written as\footnote{See our conference contribution~\cite{icc}.}
\begin{equation}\label{eq10}
\mathbf{F}_d=\mathbf{U}_d\left[(\mu_d \mathbf{I} -\mathbf{\Lambda}^{-1}_{d})^{+}\right]^{\frac{1}{2}}
\end{equation}
where $\mu_d$ is the power level for the WF algorithm similar to that in \eqref{eq5} and $\mathbf{U}_d$ is obtained through the following eigenvalue decomposition
\begin{align}\label{exp}
\mathbf{U}_d\mathbf{\Lambda}_{d} \mathbf{U}^H_d&=\mathbf{H}^H_{\perp}(\mathbf{Z} +\sigma^2 \mathbf{I})^{-1}\mathbf{H}_{\perp} \nonumber\\ 
&=(\mathbf{I}-\mathbf{U}_G\mathbf{U}^H_G)^H \mathbf{H}^H (\mathbf{Z} +\sigma^2 \mathbf{I})^{-1} \mathbf{H} (\mathbf{I}-\mathbf{U}_G\mathbf{U}^H_G).
\end{align}
It can be seen from \eqref{eq10} and \eqref{exp} that in order to obtain the CR precoding matrix, the interference-plus-noise covariance matrix $\mathbf{R}_{\rm{ur}}\triangleq\mathbf{Z}+\sigma^2_n\mathbf{I}$ needs to be estimated at the CR receiver, besides the estimation of the interference subspace $\mathbf{U}_G\mathbf{U}^H_G$ at the CR transmitter.

\subsection{Full-projection-based precoding}

To enable the estimation of $\mathbf{U}_G\mathbf{U}^H_G$ and $\mathbf{R}_{\rm{ur}}$, we propose an enhanced precoding scheme. The system diagram for this scheme is demonstrated in Fig.~2. Each CR cycle consists of sensing and transmission phases. We name the CR transmission during the primary downlink transmission as T1 and uplink transmission as T2. For T1, the space ${\mathbf{U}}_G{\mathbf{U}}^H_G$ is estimated at the CR transmitter during the primary uplink according to \eqref{eq7}--\eqref{gn} over $L_{S1}$ symbols. The estimation of $\mathbf{R}_{\rm{ur}}$ is performed at the CR receiver at the beginning of the primary downlink for a batch of $L_{S2}$ symbols via a  procedure similar to \eqref{eq7}. After obtaining these two estimates, the CR transmitter starts transmission T1 using the precoding matrix obtained by \eqref{eq10}. Then T2 follows immediately after T1 but right before the sensing phase for the next CR cycle. The CR precoding matrix for T2 can be obtained by other two sensing sessions concurrent with the sensing phase for T1. The FP-based channel projection \eqref{eq4} is employed in this proposed precoding scheme. Therefore, it is termed as FP precoding. 

It can be seen from Fig.~2 that the proposed FP precoding scheme shifts the CR cycle of the SP precoding rightwards in time. By doing this, several benefits are obtained. Firstly, introducing CR receiver sensing phases during both the primary downlink and uplink improves the CR instantaneous throughput by incorporating the interference-plus-noise covariance matrix into precoding, and consequently improves the CR throughput. Secondly, shifting the CR cycle diverts part of the CR transmission from the primary downlink to the uplink which reduces the time that primary receivers expose themselves to interference from the CR transmitter. This is beneficial to the primary network, since primary users are usually more susceptible to CR interference than the primary BS.

Theoretically, the proposed FP precoding can completely cancel the CR-primary interference if there is no error in the interference space estimation \eqref{gn}. However, the IM ability of the proposed FP precoding degrades rapidly when the CR interference-to-noise ratio, INR$\triangleq {\sigma^2_{\rm{ut}}}/\sigma^2_n$, drops below a threshold. This is due to the fact that in \eqref{gn} some components in the noise subspace may swap with those in the interference subspace when the noise amplitude $\sigma_n$ is relatively large compared to the interference channel gain\! $\sigma_{\rm{ut}}$.\! This phenomenon is\! known as\! a \!\textit{subspace swap}\footnote{The lower bound on the probability of the subspace swap has been investigated in \cite{ss_pro1} and \cite{ss_pro2}.}  \cite{subspace_swap}. 

For low INR, the interference subspace has a high probability to swap with the noise subspace. When a subspace swap happens, \eqref{exp} can be rewritten as
\vspace{-0.1cm}
\begin{align} \label{ss}
\mathbf{U}_d\mathbf{\Lambda}_{d} \mathbf{U}^H_d&\approx (\mathbf{I}-\hat{\mathbf{U}}_n\hat{\mathbf{U}}_n^H)^H \mathbf{H}^H (\mathbf{Z} +\sigma^2 \mathbf{I})^{-1} \mathbf{H} (\mathbf{I}-\hat{\mathbf{U}}_n\hat{\mathbf{U}}_n^H) \nonumber \\ 
&=\hat{\mathbf{U}}_G\hat{\mathbf{U}}_G^H \mathbf{H}^H (\mathbf{Z} +\sigma^2 \mathbf{I})^{-1} \mathbf{H} \hat{\mathbf{U}}_G\hat{\mathbf{U}}_G^H
\end{align}
\vspace{-0.1cm}
which means that the precoding matrix $\mathbf{F}_d$ and ${\mathbf{H}^H_\mathrm{ut}}$ span the same space. Therefore, when the CR INR is low the average interference power received at primary users can be expressed as
\begin{align}\label{low_inr}
I_l^{FP}=\mathbb{E}\{\mathtt{Tr} \{{\mathbf{H}^H_\mathrm{ut}} \mathbf{F}_d  {\mathbf{F}^H_d} {\mathbf{H}_\mathrm{ut}} \} \} \propto P_{cr} \sigma^2_\mathrm{ut}.
\end{align}
This suggests that the average interference power at primary users is proportional to the channel gain between CR and primary users at low CR INR.

The average CR-primary interference in the large CR INR regime becomes 
\begin{align}\label{high_inr}
I_h^{FP}&=\mathbb{E}\{\mathtt{Tr} \{{\mathbf{H}^H_\mathrm{ut}} \hat{\mathbf{U}}_d (\mu_d \mathbf{I} -\mathbf{\Lambda}^{-1}_{d})^{+} {\hat{\mathbf{U}}_d}^H \mathbf{H}_\mathrm{ut} \} \\ \label{hi1}
&= \mathbb{E}\{\mathtt{Tr} \{{\mathbf{H}^H_\mathrm{ut}} (\hat{\mathbf{U}}_d - \mathbf{U}_d) (\mu_d \mathbf{I} -\mathbf{\Lambda}^{-1}_{d})^{+} {(\hat{\mathbf{U}}_d - \mathbf{U}_d)}^H \mathbf{H}_\mathrm{ut} \}\\ \label{hi2}
&\approx \mathbb{E}\{\mathtt{Tr} \{{\mathbf{H}^H_\mathrm{ut}} (\mathbf{X}^H \mathbf{H}_\mathrm{ut})^{\dag} \mathbf{N}^H \mathbf{U}_d (\mu_d \mathbf{I} -\mathbf{\Lambda}^{-1}_{d})^{+} {\mathbf{U}_d}^H \mathbf{N} ({\mathbf{H}^H_\mathrm{ut}} \mathbf{X})^{\dag} \mathbf{H}_\mathrm{ut} \} \}\\ \label{hi3}
&=\sigma^2_n P_{cr} \mathbb{E}\{\mathtt{Tr} \{{\mathbf{H}^H_\mathrm{ut}} (\mathbf{X}^H \mathbf{H}_\mathrm{ut})^{\dag} ({\mathbf{H}^H_\mathrm{ut}} \mathbf{X})^{\dag} \mathbf{H}_\mathrm{ut} \} \}\\ \label{hi4}
&=\frac{\sigma^2_n P_{cr}}{L_{S1}}\mathtt{Tr} \{\mathbf{Q}_u \}
\end{align}
where \eqref{hi1} is due to the fact that ${\mathbf{H}^H_\mathrm{ut}} \mathbf{U}_d = \mathbf{0}$; \eqref{hi2} is obtained using the fact that $\hat{\mathbf{U}}_d - \mathbf{U}_d \approx -(\mathbf{X}^H \mathbf{H}_\mathrm{ut})^{\dag}\mathbf{N}^H \mathbf{U}_d $ for high INR \cite{per} with $\mathbf{X}\triangleq[\mathbf{x}_u(1), \mathbf{x}_u(2),\cdots, \mathbf{x}_u(L_{S1})]$, and $\mathbf{N}\triangleq[\mathbf{n}(1), \mathbf{n}(2),\cdots, \mathbf{n}(L_{S1})]$; \eqref{hi3} follows from the independence of $\mathbf{X}^H \mathbf{H}_\mathrm{ut}$ and $\mathbf{N}$ and $\mathbb{E}\{\mathbf{N}^H\mathbf{Y}\mathbf{N}\}=\sigma^2_n \mathtt{Tr}\{\mathbf{Y}\}\mathbf{I}$ for any constant matrix $\mathbf{Y}$. Note that $\mathbf{Q}_u \triangleq \mathbb{E}\{\mathbf{x_u}\mathbf{x_u}^H\}$ in \eqref{hi4} is the transmit covariance matrix for the primary user. An interesting fact can be observed from \eqref{hi4} that at high CR INR the average received interference at primary users does not depend on the interference channel ${\mathbf{H}^H_\mathrm{ut}}$. The average interference is proportional to the channel noise $\sigma^2_n$ and inversely proportional to the sensing length $L_{S1}$.

\subsection{Partial-projection-based precoding}
To further improve the throughput of the CR link, the CR transmitter 
may null its transmission to a subspace partially spanning the interference space. Therefore, we introduce another precoding scheme, namely, the PP precoding. 

The PP precoding works in a similar manner to the above proposed FP precoding except for the selection of the interference space. For the downlink CR precoding, the CR transmitter first obtains $\hat{\mathbf{\Lambda}}$ and $\hat{\mathbf{U}}$ via eigenvalue decomposition in \eqref{svd} during uplink sensing. Then, a subspace $\hat{\mathbf{U}}_m\hat{\mathbf{U}}^H_m$ partially spanning the interference space is obtained by choosing $m$ eigenvectors corresponding to the first $m$ largest eigenvalues of $\hat{\mathbf{\Lambda}}$, where $m$ can be determined according to various criteria. One candidate criterion is
\begin{equation} \label{ftd}
\frac{\sum^{M_{\rm{min}}}_{i=m+1} \lambda_i}{\sum^m_{i=1} \lambda_i} \leq r_{t/d}
\end{equation}
with $M_{\rm{min}}\triangleq\mathtt{min}(M_t, \sum^K_{k=1} M_k)$. We call $r_{t/d}$ the trivial over dominant interference ratio (TDIR). This selection process chooses $m$ dominant interference subchannels to form an estimate of the interference space and ignores the other $(M_{\rm{min}}-m)$ trivial ones. When equal power is assigned to each CR antenna, $r_{t/d}$ stands for the maximum ratio of the resulting and nullified interference to the primary receiver. Finally, substituting the estimated subspace $\hat{\mathbf{U}}_m\hat{\mathbf{U}}^H_m$ for $\hat{\mathbf{U}}_G\hat{\mathbf{U}}^H_G$ in \eqref{eq4}, the precoding matrix $\mathbf{F}_d$ for the downlink CR transmission can be obtained via \eqref{eq10}. However, we may fail to find a value of $m$ satisfying \eqref{ftd}. In this case, the proposed FP precoding is used. The uplink CR precoding can be performed in the similar manner as the downlink counterpart.

The joint probability density function (PDF) of the ordered eigenvalues $\mathbf{\lambda}\triangleq[\lambda_1,\lambda_2,\cdots,\lambda_{M_{\rm{min}}}]$ of $\hat{\mathbf{R}}_{\rm{ut}}$, with ${\lambda}_1\geq {\lambda}_2\geq \cdots \geq {\lambda}_{M_{\rm{min}}}\geq \sigma^2_n$ is \cite{jointpdf}
\begin{equation}\label{jpdf}
f_{\mathbf{\lambda}}(\lambda_1,\lambda_2,\cdots,\lambda_{M_{\rm{min}}})=\frac{1}{{P_p}^{M_{\rm{min}}}}f_{\tilde{\mathbf{\lambda}}}\left(\frac{\lambda_1-\sigma^2_n}{P_p},\frac{\lambda_2-\sigma^2_n}{P_p},\cdots,\frac{\lambda_{M_{\rm{min}}}-\sigma^2_n}{P_p}\right)
\end{equation}
where $P_p$ is the transmission power of each primary user antenna and $f_{\tilde{\mathbf{\lambda}}}(\tilde{\lambda}_1,\tilde{\lambda}_2,\cdots,\tilde{\lambda}_{M_{\rm{min}}})$ with $\tilde{\lambda}_1\geq \tilde{\lambda}_2\geq \cdots \geq \tilde{\lambda}_{M_{\rm{min}}}$ is given by
\begin{equation}\label{jpdf2}
f_{\tilde{\mathbf{\lambda}}}(\tilde{\lambda}_1,\tilde{\lambda}_2,\cdots,\tilde{\lambda}_{M_{\rm{min}}})=\frac{\prod^{M_{\rm{min}}}_{i=1}e^{-\tilde{\lambda}_i}\tilde{\lambda}^{M_{\rm{max}}-M_{\rm{min}}}_i \prod^{M_{\rm{min}}-1}_{i=1}\left[ \prod^{M_{\rm{min}}}_{j=i+1} (\tilde{\lambda}_i - \tilde{\lambda}_j)^2 \right]}{\prod^{M_{\rm{min}}}_{i=1}(M_{\rm{max}}-i)! \prod^{M_{\rm{min}}}_{i=1}(M_{\rm{min}}-i)!}
\end{equation}
with $M_{\rm{max}}\triangleq\mathtt{max}(M_t, \sum^K_{k=1} M_k)$. Therefore, the probability for the occurrence of \eqref{ftd} is
\begin{equation}\label{pm}
p_m=\int_{\mathbf{S}}f_{\mathbf{\lambda}}(\lambda_1,\lambda_2,\cdots,\lambda_{M_{\rm{min}}})\ d\lambda_1 \ \! d\lambda_2 \cdots d\lambda_{M_{\rm{min}}}
\end{equation}
where $\mathbf{S}\triangleq\{(\lambda_1,\lambda_2,\cdots,\lambda_{M_{\rm{min}}})| \ \! \eqref{ftd} \cap {\lambda}_1\geq {\lambda}_2\geq \cdots \geq {\lambda}_{M_{\rm{min}}} \geq \sigma^2_n\}$. 

In other words, for the PP precoding scheme the probabilities of using the `real' PP ($m$ satisfying \eqref{ftd} exists) and using FP are $p_m$ and $(1-p_m)$, respectively. Therefore, the CR transmitter uses $(1- p_m) \sum^K_{k=1}M_k + p_m m$ and $\sum^K_{k=1}M_k$ degrees of freedom (DoF) for interference mitigation in the PP and FP precoding schemes, respectively. Meanwhile, the DoF for CR transmission for the PP and FP precoding are $M_t-(1- p_m) \sum^K_{k=1}M_k - p_m m$ and $M_t-\sum^K_{k=1}M_k$, respectively. This means that compared to the proposed FP precoding the PP precoding scheme transfers $p_m (\sum^K_{i=1}M_k-m)$ DoF from interference mitigation to CR transmission, which leads to a higher throughput for the CR link. It can be seen from \eqref{jpdf}--\eqref{pm} that in the large INR regime, $p_m$ is fixed for a given noise power $\sigma^2_n$ and $P_p$, i.e., the probability of the `real' PP and FP does not change with the interference channel $\mathbf{H}_{\rm{ut}}^H$. Considering the fact from \eqref{hi4} that at high INRs the average interference power of FP $I^{FP}_h$ is fixed and the average interference power resulting from `real' PP $I^{PP}$ is proportional to the square of the interference channel gain $\sigma^2_{\rm{ut}}$, the overall average interference of the PP precoding $I^{PP}_h=p_mI^{PP} + (1-p_m)I^{FP}_h$ is linearly proportional to $\sigma^2_{\rm{ut}}$ for large INRs.

\section{Numerical Results \& Discussions}

In this section, the performance of the proposed precoding schemes is evaluated via simulations. We consider a scenario where a CR MIMO system coexists with a primary TDD system which has one 2-antenna BS and two single-antenna users. Each CR node is equipped with four antennas, i.e., $M_t=M_r=4$, $M_{bs}=2$, $K=2$ and $M_1=M_2=1$. The primary network works as a downlink-broadcast and an uplink multiple-access system. The primary BS uses perfect zero-forcing beamforming at both the primary downlink and uplink. The transmission power of the CR and primary networks is $1$. All the results are obtained by averaging over 2000 simulation runs.

First, we evaluate the throughput of the CR system with the proposed precoding schemes over different values of signal-to-noise ratios, SNR$\triangleq{\sigma^2_H}/{\sigma^2_n}$. In Fig.~3, the throughputs (mutual information in \eqref{eq9} averaged over a CR cycle) of the two proposed precoding schemes are compared with that of the SP precoding of \cite{ns_precoding}, \cite{practical_bf} and the P-SVD precoding with perfect CSI of \cite{rui}. The system setup is as follows: $L_{S1}=L_{S2}=L_{T2}=50,\ L_{T1}=350,\ \sigma^2_H=\sigma^2_{\rm{ut}}=1,\ P_{cr}=1$, and $r_{t/d}=0.1$. It can be seen that the proposed FP/PP precoding schemes lead to higher CR throughput than the SP precoding, and the throughput gain becomes larger as the SNR increases.

Fig.~4 evaluates the impact of CR INR on the CR throughput and the resulting CR-primary interference under different precoding schemes.
It has the same setup as that of Fig.~3 with $\sigma^2_n=10^{-4}$. By comparing Fig.~4(a) with Fig.~4(b), it can be seen that
the proposed FP/PP precoding schemes outperform the SP counterpart at low INRs, since they lead to higher CR throughput without introducing extra interference. At high INRs, both the proposed FP and SP precoding schemes have fixed interference, and there is a fairly good agreement between the derived and simulated interference of the FP precoding. Another phenomenon which can be seen from Fig.~4(b) is that the interference of the SP precoding is slightly smaller than that of the FP precoding. This is due to the fact that the sensing of the SP precoding is longer than the uplink sensing of the FP precoding. Moreover, at high INRs the interference of the proposed PP precoding is linearly proportional to the CR INR, which supports our analysis in Section~IV.B.

\section{Conclusions}

In this paper, two SP-based precoding schemes, namely, FP and PP precoding, have been proposed for CR MIMO systems to mitigate the interference to the primary network and improve the CR throughput. These two precoding schemes are capable of estimating the CSI of interference channels between primary and CR networks and can account for the interference from the primary system via a novel sensing approach. Therefore, no extra signaling is required between primary and CR systems, which consequently eases the deployment of CR networks. The performance of the proposed precoding schemes has been evaluated and compared to that of the existing precoding approaches. It has been demonstrated that the FP precoding can boost the CR throughput and does not introduce extra interference to the primary system in the low INR regime. The PP precoding can further improve the CR throughput if the primary system can tolerate some extra interference.
\vspace{-0.2cm}
\begin{center}
\textsc{{Acknowledgments}}\\
\end{center}
\vspace{-0.3cm}
{\small Z. Chen, C.-X. Wang, X. Hong, and J. Thompson acknowledge the support from the Scottish Funding Council for the Joint Research Institute in Signal and Image Processing between the University of Edinburgh and Heriot-Watt University, as a part of the Edinburgh Research Partnership in Engineering and Mathematics (ERPem). S. Vorobyov acknowledges the support in part from the Natural Sciences and Engineering Research Council (NSERC) of Canada and in part from the Alberta Ingenuity Foundation, Alberta, Canada. F. Zhao acknowledges the support from National Natural Science Foundation of China (NSFC) (Grant No.: 60872022). C.-X. Wang and F. Zhao acknowledge the support of the Key Laboratory of Cognitive Radio and Information Processing (Guilin University of Electronic Technology), Ministry of Education, China. X. Ge acknowledges the support from the NSFC (Grant No.: 60872007), National 863 High Technology Program of China (Grant No.: 2009AA01Z239), and the Ministry of Science and Technology (MOST) of China, International Science and Technology Collaboration Program (Grant No.: 0903). The authors acknowledge the support from the RCUK for the UK-China Science Bridges Project: R\&D on (B)4G Wireless Mobile Communications.}


\newpage
\begin{center}
\small {\ \ \ }
\end{center}

\vspace{1cm}
\begin{center}
\begin{tabular}{c}
\hskip -2.5cm\epsfxsize=9cm\epsffile{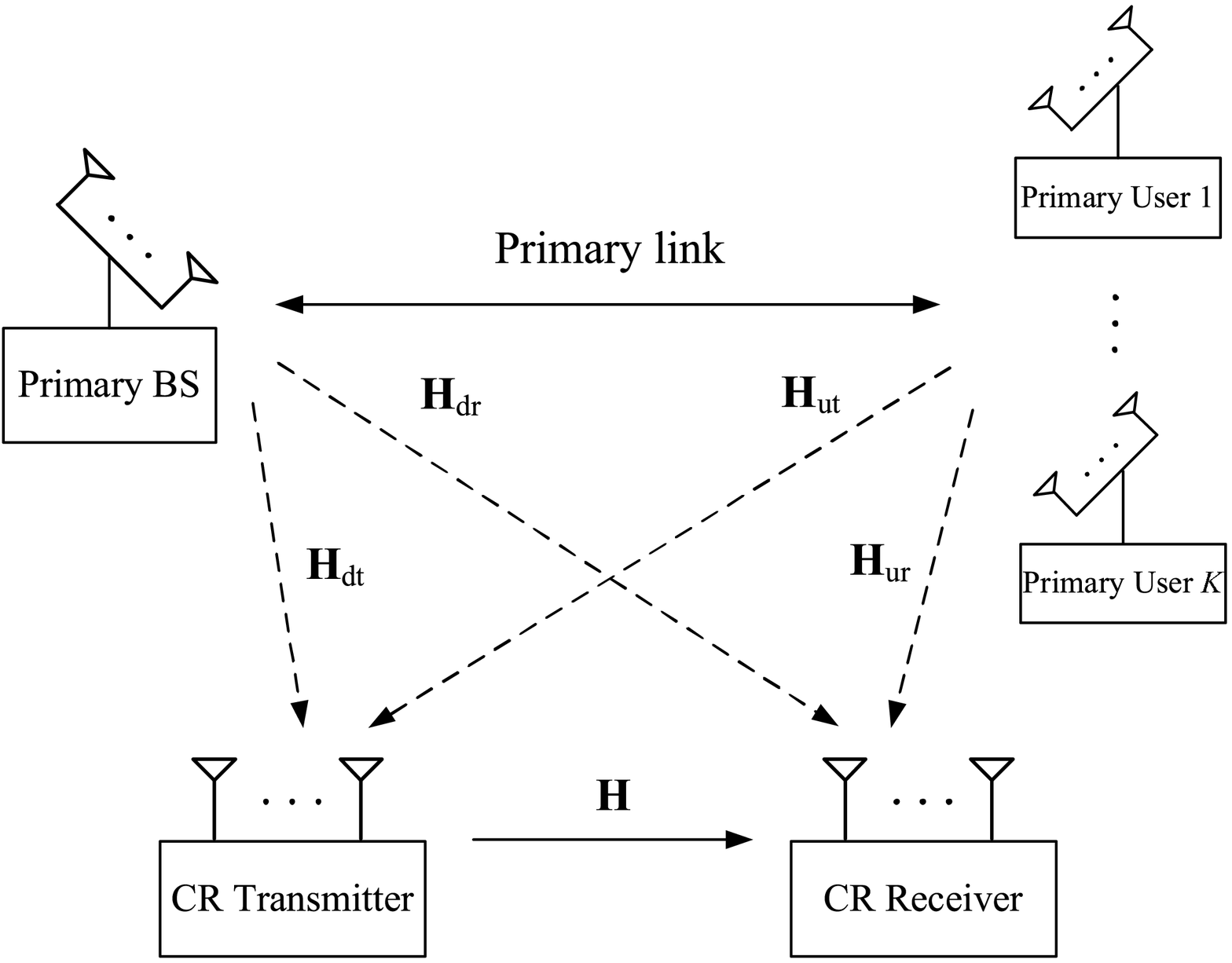} \\
\end{tabular}\\
\vspace*{-0.2cm}
\begin{center}
\small Fig.~1. A CR MIMO transmitter-receiver pair coexists with a primary TDD system.
\end{center}
\end{center}

\vspace{4.2cm}
\begin{center}
\begin{tabular}{c}
\hskip -3cm\epsfxsize=12.5cm\epsffile{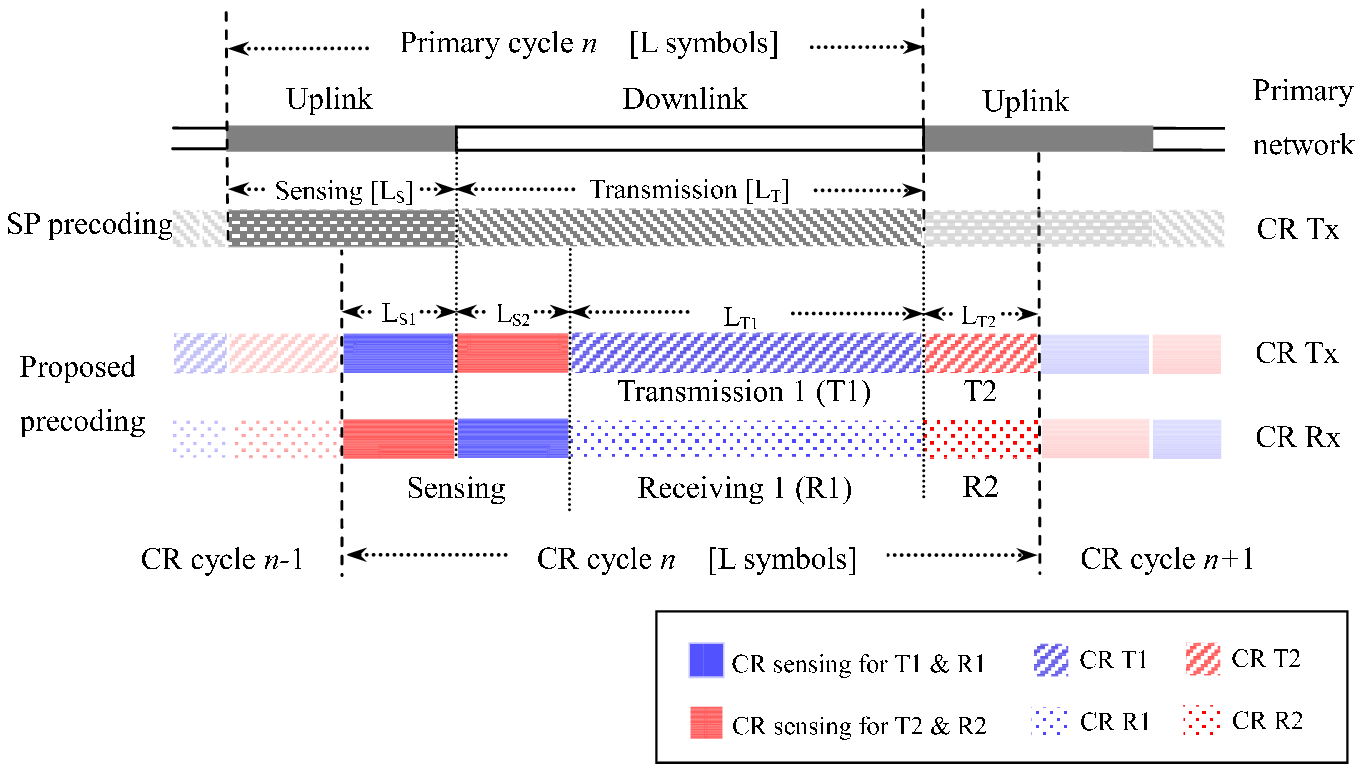} \\
\end{tabular}\\
\vspace*{-1.7cm}
\begin{center}
\small Fig.~2. System diagram for the proposed precoding schemes.
\end{center}
\end{center}

\begin{center}
\begin{tabular}{c}
\hskip -1cm\epsfxsize=12cm\epsffile{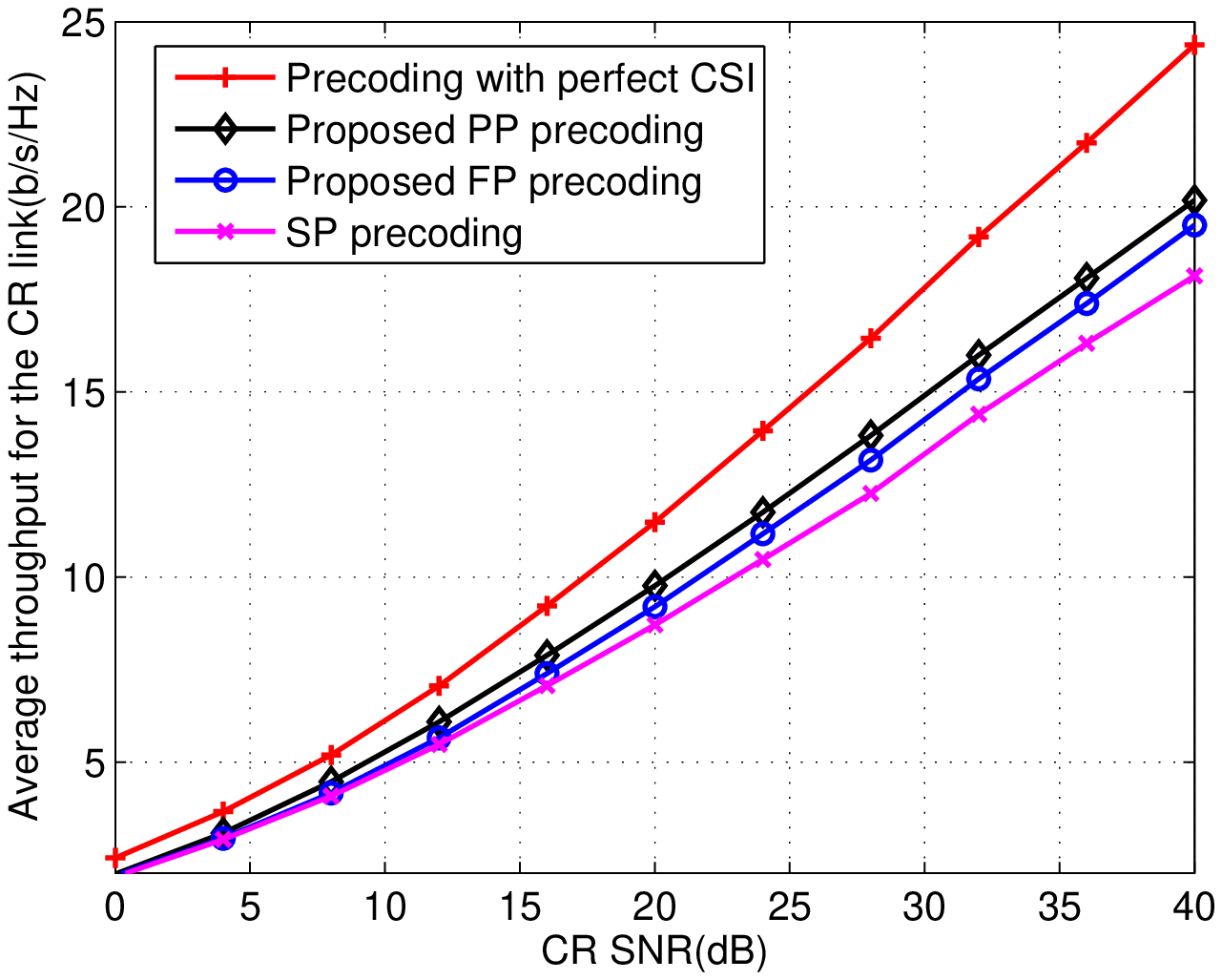} \\
\end{tabular}\\
\vspace*{-0.2cm}
\begin{center}
\small Fig.~3. CR throughput under different precoding schemes ($M_t=M_r=4,$ $M_{\rm{bs}}=2,$ $K=2,$ $M_1=M_2=1,$ $L_{S1}=L_{S2}=L_{T2}=50,$ $L_{T1}=350,$ $\sigma^2_H=\sigma^2_{\rm{ut}}=1,$ $P_{cr}=1,$ and $r_{t/d}=0.1$).
\end{center}
\end{center}

\vspace{1cm}
\begin{center}
\begin{tabular}{c}
\hskip -1.5cm \epsfxsize=20cm\epsffile{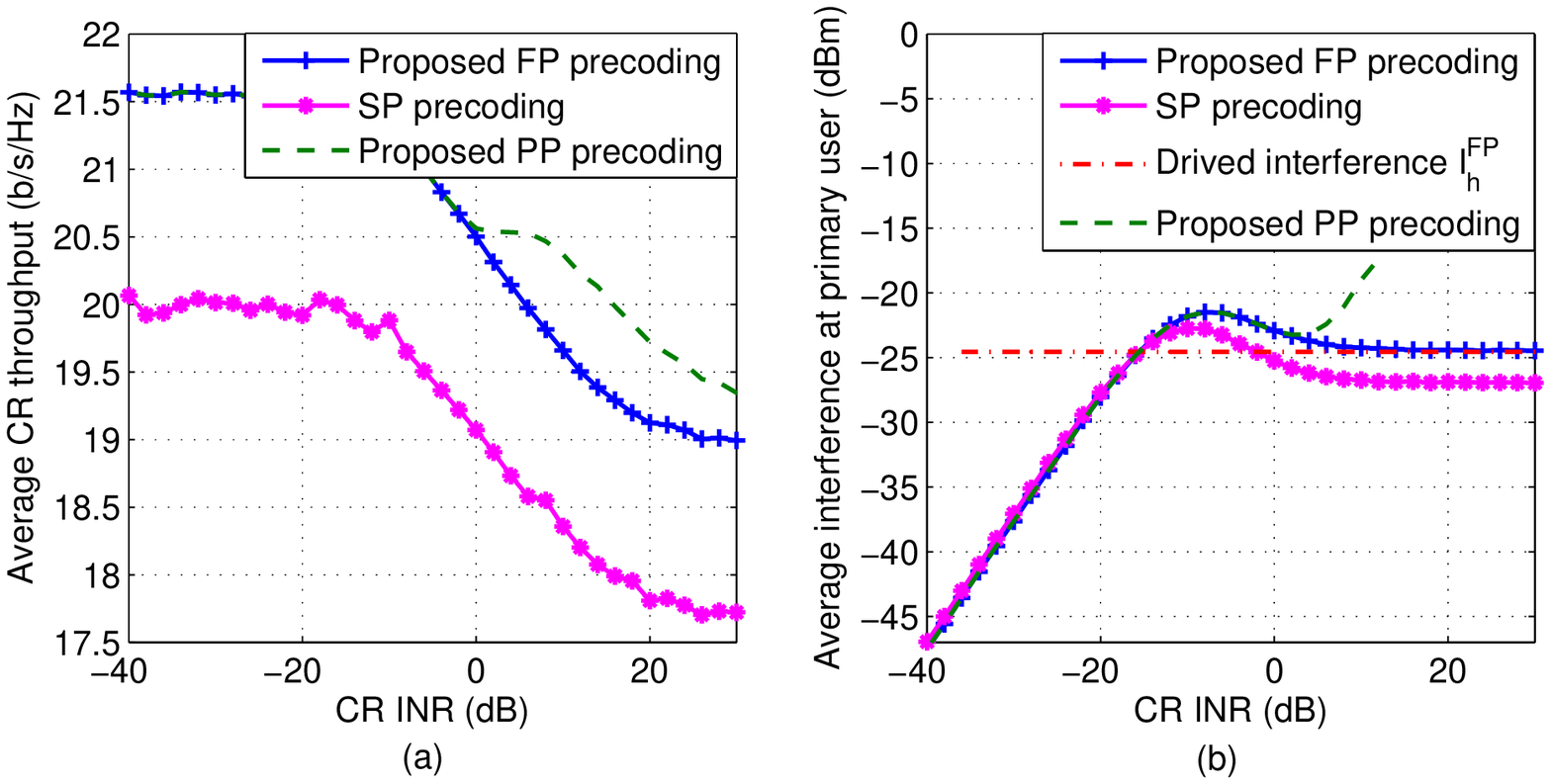} \\
\end{tabular}\\
\vspace*{-0.2cm}
\begin{center}
\small Fig.~4. (a) CR throughput and (b) resulting interference of different precoding schemes ($M_t=M_r=4,$ $M_{\rm{bs}}=2,$ $K=2,$ $M_1=M_2=1,$, $L_S=100,$  $L_{S1}=L_{S2}=L_{T2}=50,$ $L_{T1}=350,$ $\sigma^2_H=1,$ $P_{cr}=1,$ $r_{t/d}=0.1$, and $\sigma^2_n=10^{-4}$).
\end{center}
\end{center}
\end{document}